# Anti-polar state in BiFeO$_3$/NdFeO$_3$ superlattices

M. A. Khaled[1], D. C. Arnold[2], B. Dkhil[3], M. Jouiad[1], K. Hoummada[4], M. El Marssi[1], H. Bouyanfif[1,*]

[1] LPMC UR2081, Université de Picardie Jules Verne, 33 Rue Saint Leu, 80000 Amiens, France

[2] School of Physical Sciences, University of Kent, Canterbury, Kent, CT2 7NH, UK

[3] Université Paris-Saclay, CentraleSupélec, CNRS UMR8580, Laboratoire Structures, Propriétés et Modélisation des Solides (SPMS), 91190 Gif-sur-Yvette, France

[4] Aix Marseille Univ, CNRS, Université de Toulon, IM2NP, 13397, Marseille, France

**Abstract**

Antiferroelectrics are promising materials for high energy density capacitors and the search for environmentally-friendly and efficient systems is actively pursued. An elegant strategy to create and design new (anti)ferroic system relies on the use of nanoscale superlattices. We report here the use of such strategy and the fabrication of nanoscale BiFeO$_3$/NdFeO$_3$ superlattices and in depth characterization using high resolution X-ray diffraction and Transmission Electron Microscopy. The structural analysis at atomic scale demonstrates that such superlattices host anti-polar ordering most likely described by an antiferroelectric-like Pbnm symmetry. Temperature dependence of anti-polar state and structural transition further hint that the stability of the anti-polar state is controlled by the BiFeO$_3$ layer thickness within the stacking and, in a more moderate way, by interlayer strain. Discovery of such polar arrangement in superlattices and the possible generalization to the whole rare-earth family pave the way to new platforms for energy storage application as well as nano-electronic devices.

*Corresponding author: houssny.bouyanfif@u-picardie.fr

**Introduction**

Antiferroelectric (AFE) materials characterized by a double-hysteresis polarization versus electric field loop have recently triggered intensive studies due to their application potential in energy storage technologies [1,2]. Due to environmental issues with the mostly investigated AFE system i.e. PbZrO$_3$, design of lead free AFEs is urgently required and Bi-based systems such as BiFeO3-based AFEs have attracted a lot of attention. BiFeO3 (BFO) is actually the most studied multiferroic material due to its outstanding properties [3]. BFO crystallizes with rhombohedral R3c (no.161 and pseudo-cubic lattice parameter of 3,96 Å) symmetry and exhibits a robust ferroelectricity with a relatively high Curie Temperature ($T_c$ = 1100 K) and a polarization value up to ~90 µC.cm$^{-2}$ [3,4]. BFO adopts an ($a^-a^-a^-$) oxygen octahedra tilt/rotation system in Glazer's notation [5], and polar cation displacements along the [111] axis [6]. At $T_c$, a first-order phase transition from ferroelectric to paraelectric order occurs upon heating. A GdFeO3-like orthorhombic Pbnm (no.62; equivalent to the Pnma space group but with switched setting axis) symmetry with ($a^-a^-c^+$) tilt/rotation system was found for the high-temperature phase [7]. The partially filled 3d shells of $Fe^{+3}$ cations are responsible for the room-temperature magnetism in BFO. A canted G-type antiferromagnetic order is observed below $T_N$ = 640K with a spiral modulation of magnetic spins preventing the development of a net magnetization and only a poor magnetoelectric coupling is achieved [8].

Site-engineering by substituting BFO cations with other chemical elements was proposed as a solution to enhance physical properties [9]. $Fe^{3+}$-site cations substitution with Mn or/and Ti elements was found to reduce the leakage current [10] and partial substitution of $Bi^{3+}$ cations has also been widely addressed. Particular attention has been paid to introducing rare earth (RE) elements at the Bi-site and complex phase diagrams were found for $La^{3+}$, $Nd^{3+}$, $Sm^{3+}$, and $Dy^{3+}$ [11-13]. Most of the previous investigations revealed an emergent intermediate phase dependent on the RE element concentration in the BFO solid solutions. Such an effect was correlated to a so-called Morphotropic Phase Boundary (MPB) between the two ground-state phases (i.e. rhombohedral and orthorhombic phases of BiFeO3 and RE-FeO3, respectively) [12,14]. This MPB, that is more a region than a boundary, appears in a narrow RE concentration range depending on the ionic radius of the RE element [12]. Interestingly, in this MPB region, enhanced piezoelectric response was reported and attributed to the electromechanical softening induced by the competing structures [15]. Moreover, local characterizations using Transmission Electron Microscopy (TEM) have revealed a stable PbZrO3-like antipolar distortion [11,14] within the MPB. A ferromagnetic order was also observed there for $Nd^{3+}$-substituted BFO along with evidence of magnetoelastic coupling [16].

The richness of the physics and the plethora of remarkable properties related to the existence of the MPB have motivated multiple investigations on epitaxial thin films and heterostructures. Solid solution thin films were synthesized using combinatorial pulsed laser deposition (PLD) and different features were observed. Antiferroelectric-like double-hysteresis loops were observed close to the MPB for the case of RE elements with smaller ionic radius ($Sm^{3+}$, and $Dy^{3+}$), while an ordinary ferroelectric to paraelectric crossover was found for higher ionic radius e.g. $La^{3+}$ [12].

Artificially modulated heterostructures such as superlattices (SLs) are also employed to better understand the competing interactions that exist within the MPB. For instance, $BiFeO_3$/$LaFeO_3$ (BFO/LFO) SLs deposited on different substrates have been studied by different groups [17-21]. Interestingly, a strain- and symmetry-mismatch-induced stabilization of the AFE-like state was reported in BFO/LFO SLs [18]. This AFE-like state shows a G-type antiferromagnetic state similar to the bulk BFO system but is not observed when deposited on (111)-oriented $SrTiO_3$ substrate [19,20]. Moreover, $BiFeO_3$/$Bi_{(1-x)}Sm_xFeO_3$ SLs have been also fabricated using combinatorial PLD [22] and evidence of an AFE-like state and enhanced dielectric properties are reported [23]. The exact origin of the AFE-like state in BFO based SLs clearly needs in depth analysis (electrostatic confinement, strain and/or symmetry mismatch) considering the potential application of lead free AFE systems in energy storage with high energy density efficiency.

In this work, to further understand and generalize the route for the emergence of the AFE-like states in $BFO/REFeO_3$ SLs we choose to investigate $BiFeO_3$/$NdFeO_3$ (BFO/NFO) SLs. The smaller pseudo-cubic lattice parameter (3.89 Å) of NFO compared to that of LFO (3.92 Å) would generate more interfacial strain in the BFO/NFO SLs, compared to BFO/LFO SLs [18] knowing BFO has a pseudo-cubic parameter of 3.96 Å. Influence of interlayer strain on the emergence of AFE-like states can be thus explored here. We report below the successful growth of $(BFO_{(1-x)\Lambda}/NFO_{x\Lambda})\times 20$ SLs with a varied ratio x of NFO within the period "$\Lambda$". The total number of period $\Lambda$ (20) is kept constant. Structural characterizations were carried out on 5 samples with fixed bilayer/period thickness but different ratios of BFO and NFO to understand the structural competition at the heterointerfaces and to confirm and generalize the emergence of AFE-like features in BFO/RE-FO SLs.

**Experimental details**

BFO/NFO superlattices were grown directly on (001)-oriented $SrTiO_3$ substrates by Pulsed Laser Deposition (PLD) using a KrF excimer laser (wavelength 248nm). A stoichiometric homemade ceramic target of $NdFeO_3$ was used whereas an extra and usual amount of 10% Bi was introduced to the $BiFeO_3$ target to compensate bismuth volatility. Additionally, 5% of Mn is added to the same BFO target to reduce leakage current [10]. The growth conditions are shown in Table S1 and Table S2 presents the characteristics of the SLs with a schematic describing the stacking of the layers. High-resolution X-ray Diffraction (HR-XRD) were performed using a four circle Bruker Discover D8 diffractometer (monochromatic beam K$_{\alpha 1}$ (Cu) = 1.54056 Å using a double bounce Ge (220) hybrid monochromator). High-Temperature XRD measurements were performed using an Anton Paar DHS 1100 stage oven up to 700°C. Electron microscopy cross-sectional analysis was performed on lamella prepared by Focused Gallium Ion Beam (FIB). The lamella thicknesses were about 100 nm obtained by milling using a Thermo Fisher dual-beam HELIOS 600 nanolab setup. High-resolution transmission electron microscopy (HRTEM) images were collected with a Field Emission Gun (FEG) Titan microscope by FEI operated at 200 kV. The microscope was equipped with a spherical aberration correction (Cs) system and a contrast diaphragm of 60 μm.

**Results and discussions**

Figure 1 (a) presents the room temperature θ-2θ diffraction patterns of the 5 SLs over a large range of 2θ angles. No parasitic phases are detected (within the resolution limit) and only SLs related Bragg peaks (satellites) are observed in proximity to the STO *(00l)* diffraction peaks. These regularly spaced satellite peaks are proof of a periodic chemical modulation along the out-of-plane growth direction [19]. It is not possible to assign a peculiar diffraction peak for BFO layers and one for the NFO layers. The superlattice diffracts as a whole and the resulting diffraction pattern is a convoluted product of the superlattice periodicity and factor structure of both BFO and NFO layers (see ref [24]). The deduced period from the angular distance between two consecutive satellites is in perfect agreement with the desired 6.5 nm period (bilayer thickness) and a total thickness of 130 nm is estimated for the SLs (20 bilayers/periods). Considering the ~3.9 Å pseudo-cubic lattice parameter of our materials, each period consists of about 16 unit cells (+/-1u.c., given the accuracy of our techniques) distributed between BFO and NFO.

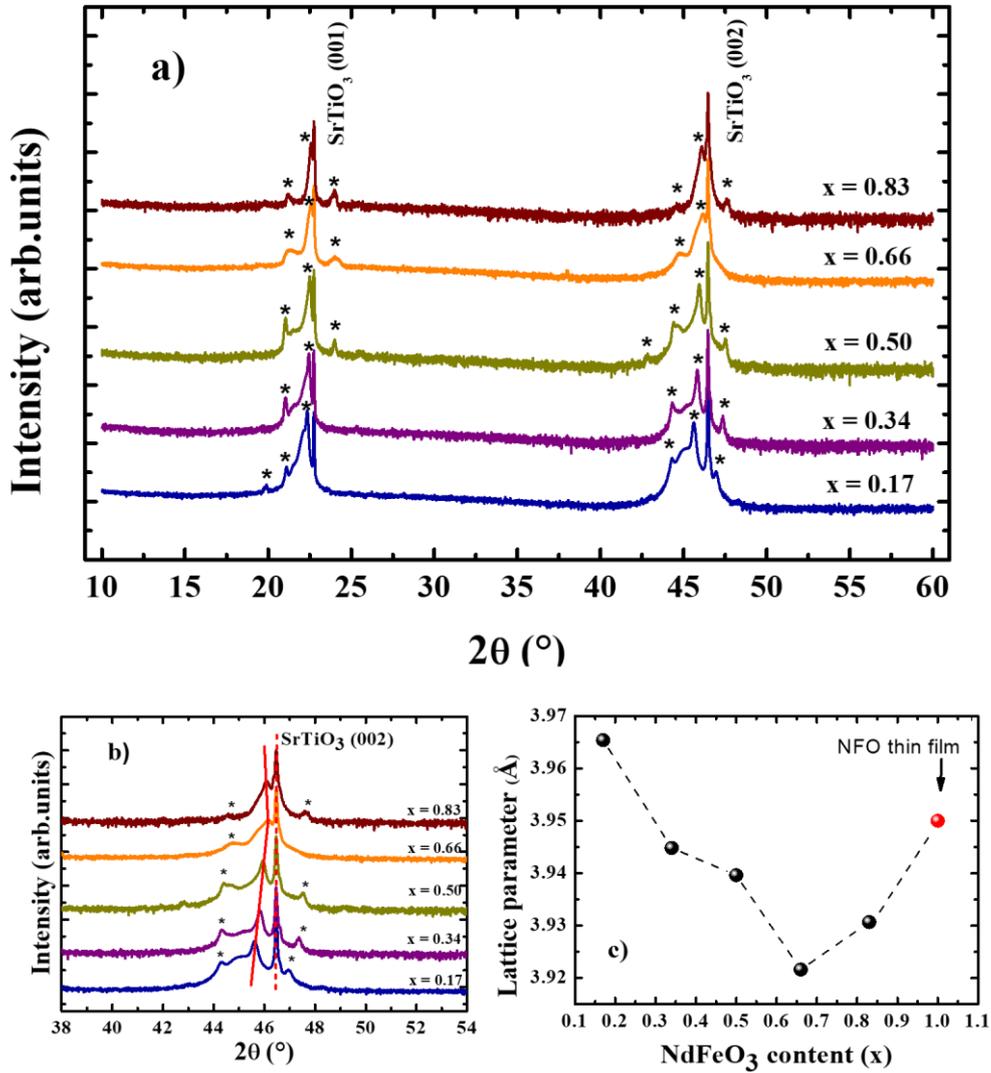

Figure 1. X-ray diffraction analysis of the BFO/NFO SLs. a) θ-2θ diffraction patterns of SLs with different NFO content x in the period. Asterisks (*) indicates SL satellite peaks b) Zoomed in section around the (002) STO diffraction peak. Continuous and dashed red lines are guides for the eyes of the most intense SLs satellite peak and the STO diffraction peak, respectively. c) Average out-of-plane lattice parameter versus NFO content x. NFO thin film (x = 1) value (red dot) is shown for comparison purpose.

SLs satellite peaks are well defined (see asterisks) and zoom in figure 1 (b) shows a dependence of the most intense satellite peak with the ratio x. A broad contribution is also observed for SLs with x below 0.66 and disappears for x = 0.83. This behavior is probably due to the interlayer strain due to the different lattice parameters of BFO and NFO; NFO (BFO) being under tensile (compressive) strain if pseudo-cubic lattice parameters are considered (3.89Å for NFO versus 3.96 Å for BFO). Strain gradients and twin patterns as observed by HRTEM (see figure 3) may explain this second contribution for rich BFO SLs. Using Bragg's law, an average lattice parameter can be estimated from the angular position of the principal satellite peak. This lattice parameter is directly correlated to the out-of-plane lattice parameters of BFO and NFO constitutive layers and will be therefore impacted by any structural changes within the SLs. The obtained values of the average lattice parameters are given in figure 1 (c). A decrease of the average lattice parameter is observed when x increases up to x = 0.66. Then the average lattice parameter raises up and extrapolates to the NFO thin film out-of-plane lattice parameter (a 125 nm thick film) on further increase of NFO ratio x. Interestingly, a similar evolution of the average out-of-plane lattice parameter has been observed in BFO/LFO SLs [18]. Such structural change has been explained as a diminution of the anti-polar distortion on increasing LFO content in the period. A similar structural behaviour might be taking place in the BFO/NFO SLs considering the very close proximity between $(Bi,La)FeO_3$ and $(Bi,Nd)FeO_3$ bulk systems. The observed minimum at a particular NFO content suggests a structural change in the SLs driven by the modification of NFO ratio x. In order to better understand the structure of the SLs, rocking curve measurements (ω-scans) and reciprocal space mapping (RSM) have been performed and are presented in figure 2.

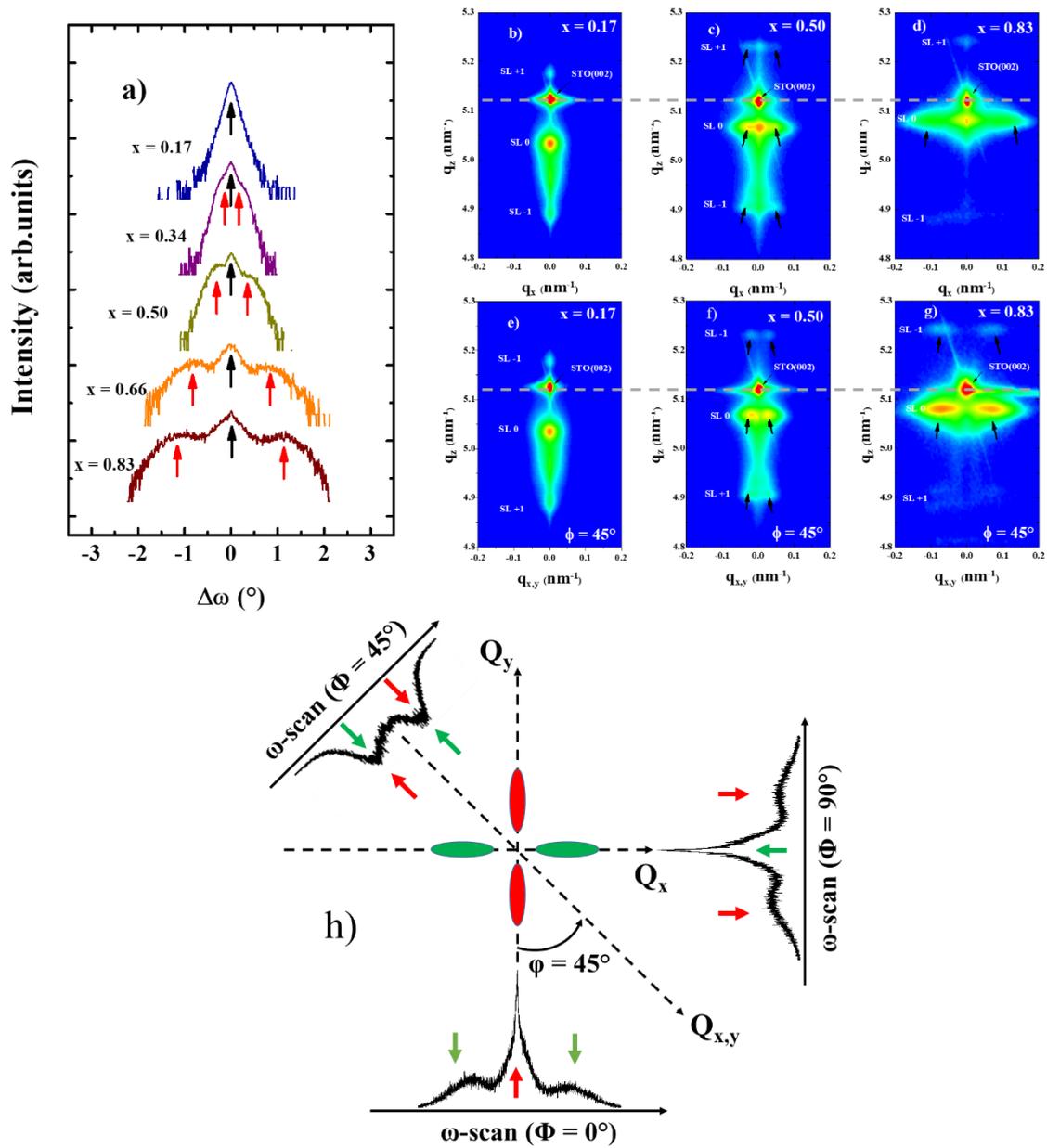

Figure 2. a) Rocking curves measured on the most intense SL satellite peak next to the STO (002) reflection (X-ray beam parallel to [100] STO direction or equivalently $\Phi = 0°$ azimuth). Black and red arrows respectively indicate the central peak and reflection due to in-plane ordering. b) to d) Reciprocal space mapping around STO (002) for samples with three different NFO content at $\Phi = 0°$ azimuth. e) to g) Reciprocal space mapping around STO (002) for samples with three different NFO content at $\Phi = 45°$ azimuth. Reflections/satellites due to Inplane ordering are shown by black arrows for rich NFO SLs. h) Distribution of diffracted intensity in reciprocal space due to in plane domains and associated rocking curve observed for 3 different azimuths ($\Phi = 0°, 45°$, and $90°$) for BFO-poor SLs.

A clear trend is evidenced in the evolution of the rocking curve measured on the most intense SL satellite peaks as presented in figure 2 (a). For the BFO-rich SL (low x), a single main peak is observed while satellites (red arrows) aside the central peak (black arrow) progressively emerge on increasing x. Moreover, the angular distance between the aside satellites and the central peak is shown to increase when x increases. Generally, regularly spaced satellites in rocking curves are due to periodic in-plane structural domains and/or ferroelectric (180° ferroelectric domains or polar exotic textures) ordering [25,26]. To understand and identify the origin (ferroelastic and/or ferroelectric) of these in-plane satellites, RSM around the (002) STO family of planes have been collected and are presented in figure 2 (b-g). RSMs were collected at different azimuths (only $\Phi = 0°$ and 45° are shown) and different observations can be made on the RSM. Due to the chemical modulation along the out-of-plane growth direction, satellite peaks are observed along the $q_z$ direction and denoted as SL-1, SL0, and SL+1. Moreover, the additional in-plane satellites detected on the rocking curves (from the SL0 most intense peak) are observed along $q_x$ and indicated by black arrows in figure 2 (c) and (d). Such in-plane satellite peaks are also convoluted with the out-of-plane chemical modulation and are also observed on the SL-1 and SL+1 peaks (with in-plane $q_x \neq 0$ and $q_{xy} \neq 0$) highlighting the good structural quality of our samples. The out-of-plane coherence length of such domains can be estimated using Debye-Scherrer law and obtained values are close to the total thickness of the SLs. The domains responsible for such reflections are therefore coherent through the whole stacking and the associated ordering exists both in BFO and NFO layers. Another important signature of these structural features is characterized by the disappearance of the central peak for azimuths 45° away from the high symmetry in-plane direction (RSM figure 2 (b-d) to compare with (e-g)). This can be explained by a distribution of the diffracted intensity in the reciprocal space as depicted in figure 2 (h). Such anisotropic distribution of reflections hints at ferroelastic domains (ferroelectric domain walls are often quasi isotropic and do not depend on the azimuth). The fact that BFO-poor SLs present strong reflections tilted away from $q_x = 0$ may be connected to the orthorhombic twins typically observed in the Pbnm orthoferrite system like NFO. In order to have access to the in-plane structural component, RSMs have been collected around the asymmetric (103) reflection of the STO substrates. Figure 3 compares RSM around the (103) STO family of planes and shows a quasi-tetragonal structure (nodes aligned in $q_x$ with STO Bragg reflection) for the BFO-rich SL while multiple reflections are detected for BFO-poor SLs.

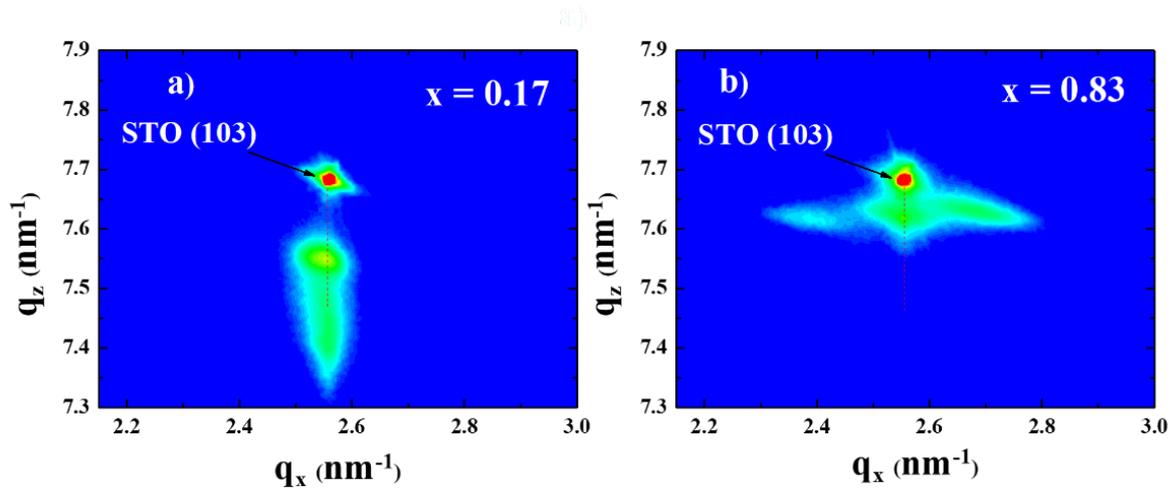

Figure 3. Reciprocal space mapping around (103) STO family of plane for a) x = 0.17 (BFO-rich SL) and b) x = 0.83 (BFO-poor SL).

The observed diffraction pattern for BFO-poor SLs is close to tensile strained of pure NFO thin films grown on STO substrates (see supplementary information showing orthorhombic twins). We therefore infer a ferroelastic origin of the periodic in plane structure on the rocking curves and (002) RSM. The increased angular distance between satellites on the rocking curves may thus be related to the increased NFO layer thickness in the SLs period. Such a feature could be explained by a progressive trend toward a bulk-like Pbnm distortion as seen on NFO thin films (figure S1). The above results suggest that SLs sustain, similarly to NFO thin films, a Pbnm orthorhombic-like twining for $0.17 < x \leq 0.34$ despite the structural change at x = 0.66 (on the average out-of-plane lattice parameter). A Pbnm-like structure for the whole set of SLs would explain such observation with a structural/polar change occurring only within the BFO layers (e.g. anti-polar to paraelectric for x above 0.66). We note that the existence of an AFE Pbnm state has been already discussed in both BFO-based solid solutions and SLs [1,21] (a Pnma setting equivalent to Pbnm is chosen in these two references). Access to the local structure, especially of BFO layer, is clearly required and a High-resolution Transmission Electron Microscopy (HRTEM) investigation has been used to probe the domain structure within the SLs. Figure 4 presents HRTEM analysis of the SL with a nominal x = 0.17 content of NFO in the period i.e. the BFO-rich SL.

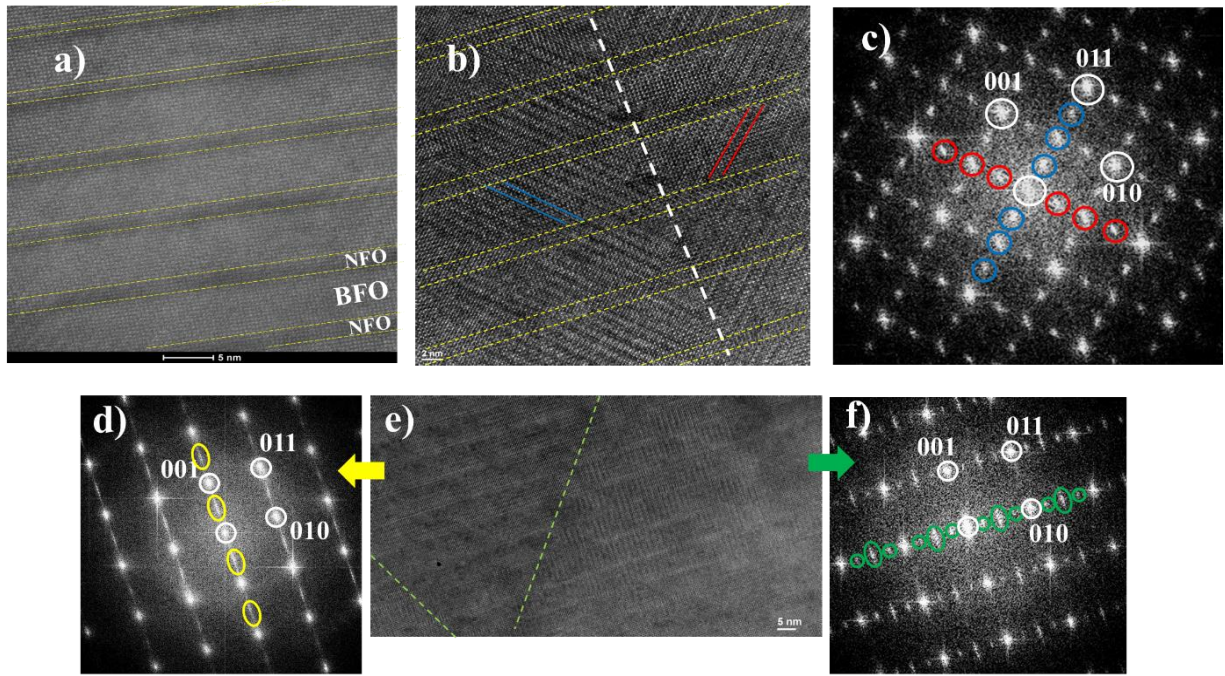

Figure 4. a) Scanning Transmission Electron Microscopy (STEM) image using a High Angle Annular Dark Field (HAADF) detector taken for an SL with x = 0.17 (BFO-rich SL). b) HRTEM image for the same sample with two PbZrO$_3$-like ordering (nano-lamellae at 45°) and vertical domain wall. c) associated FFT (Fast Fourier Transform) image highlighting PbZrO$_3$-like ¼{011} reflections. d) FFT of the left zone in e) showing ½[001] reflections due to the pseudo-cubic unit cell doubling along the out-of-plane direction. e) HRTEM image showing another zone of the SLs with two different twin variants with unit cell doubling along the out of plane direction (left zone) and in-plane quadrupling of the unit cell respectively. f) shows the FFT of the right part of e) with ½[010] and ¼[010] reflections. All images are taken at the (100) zone axis.

The HAADF-STEM image (figure 4 (a)) confirms the thicknesses of the BFO (5.2 nm) and NFO (1.3 nm) layers imposed during the growth process and evaluated by XRD analysis. Similarly to BFO/LFO SLs [18], we evidence ultra-thin lamellae tilted at 45° relative to substrate plane and two nanostructures are shown separated by a vertical domain wall, as shown in figure 4 (b). These 45° lamellae are confined within the BFO layers but a continuity can be seen from one BFO layer to another. A compatibility of the BFO and NFO lattices is therefore inferred and further confirmed by the Fast Fourier Transform (FFT) image shown in figure 4 (c). Indeed, some of the PbZrO$_3$-like ¼{110} reflections are modulated along the growth direction by the chemical modulation. Note that the vertical domain wall viewed in figure 4 (b) if compared with those observed in PbZrO$_3$ is a 90° domains wall (angle between the equivalent polar Néel vectors of the two domains). Importantly, we demonstrate here the observation of an AFE PbZrO$_3$-like structure within the BFO layers, and a clear structural compatibility is evidenced with the NFO layers. Figure 4 (e) and the corresponding FFTs (figure 4 (d) and (f)) also highlight the rich variety of twin variants within the SL. Indeed, the right part of the figure 4 (e) is associated with a domain with the in-plane PbZrO$_3$-like c$_o$ axis along the [010] direction and parallel to the NFO c$_o$ long axis (unit cell doubling along this direction). The FFT (figure 4 (f)) confirms this in-plane orientation with strong ½[010] but also weak ¼[010] reflections reminiscent of the AFE NaNbO$_3$-like structure (a$^-$a$^-$c$^+$/a$^-$a$^-$c$^-$). On the left side of figure 4 (e) only indication of the NFO orientation is clear with the orthorhombic c$_o$ long axis along the out-of-plane growth direction (see FFT figure 4 (d) with ½[001] reflections). The observed twinning pattern and associated distortions are maybe too weak to be detected by conventional XRD and only a single peak is observed on the rocking curves for the x = 0.17 SL. As a possible symmetry candidate of the anti-polar BFO/NFO structure we note that an AFE Pbnm-like symmetry is proposed in BFO/LFO SLs with features very close to our experimental observations [21]. Such phase were found to be lower in energy than both regular Pbnm and PbZrO$_3$-like Pbam symmetries and characterized by complex oxygen octahedra tilting and cation displacements. Beyond energy storage application, such complex structure with potential softness may be also interesting in devices where the negative capacitance is exploited [26]. This AFE Pbnm-like symmetry is also discussed in the context of nanoscale local ordering in (Bi,Nd)FeO$_3$ solid solutions [1]. We would like to bring to the attention of the reader that similar anti-polar structures were also observed in BFO/(La,Sr)MnO$_3$ SLs [27]. In this latter structure, BFO is combined with a metal and a screened/weaker depolarizing field is more likely expected inside the BFO layers. To better understand the SLs phase stability, we have performed a temperature dependent XRD analysis.

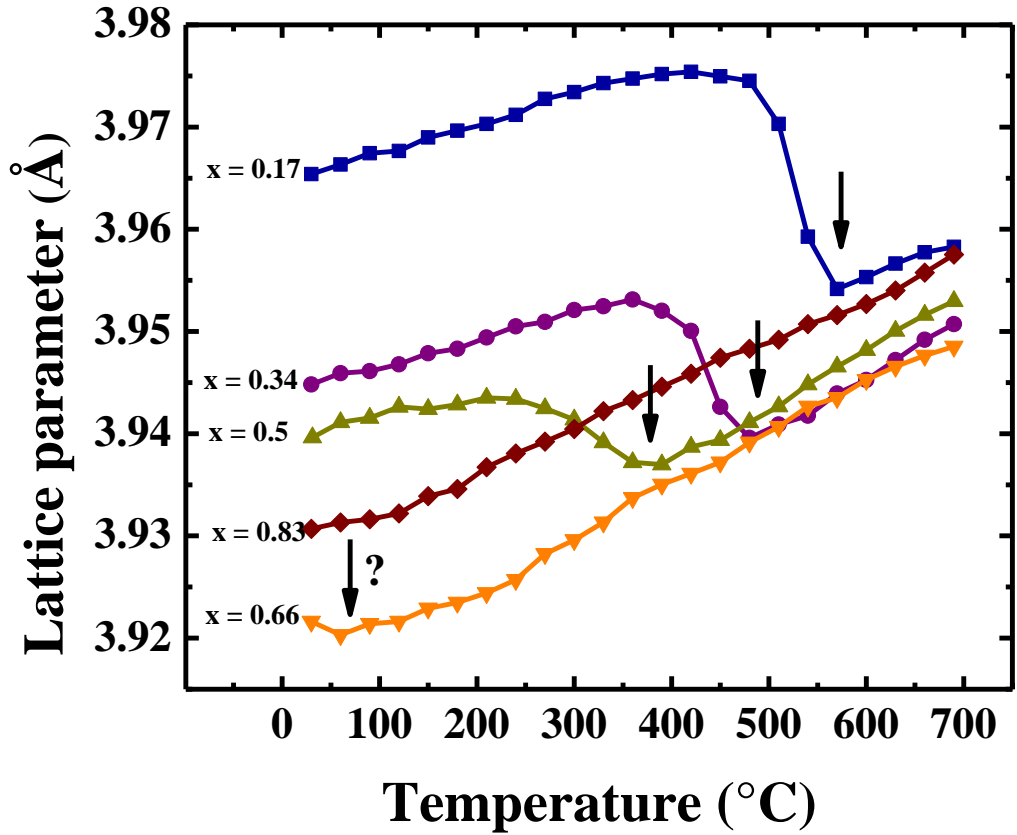

Figure 5. Average out-of-plane lattice parameters versus temperature for the different SLs. Black arrows indicate structural anomalies and the BFO-poor SL (x=0.66) deserves more investigation on the below room temperature range to confirm a possible structural anomaly.

The method to study the structural behaviour of the SLs consists of collecting θ-2θ diffraction patterns at different temperatures and exploring the changes observed on the SLs satellite peaks (Figure S2). To better quantify such changes, average lattice parameters versus temperature are plotted for all SLs in figure 5 (a). Along the linear increase of the lattice parameters (thermal expansion) we clearly see abrupt changes at a critical temperature that is strongly dependent on the NFO content x (critical anomalies are shown with black arrows). The higher the NFO content x the lower the critical temperature. In fact, the critical temperature seems to scale with the BFO thickness as also observed in BFO/LFO SLs [18] for which similar behaviours have been seen. The structural changes evidenced in figure 5 are therefore interpreted as a structural phase transition from a low temperature anti-polar phase to a high-temperature paraelectric phase. The reproducibility of such transition and the absence of modifications of the XRD θ-2θ patterns after thermal cycling exclude any temperature induced strong interdiffusion or decomposition. The high-temperature paraelectric phase is naturally described by the Pbnm symmetry adopted by both bulk BFO and NFO as well as all orthoferrites at sufficiently high temperatures. The SL with the higher NFO content does not however show abrupt anomalies but only smooth changes (within the resolution limit and temperature range). This could suggests a close proximity to the Pbnm paraelectric symmetry already discussed above for the room temperature XRD analysis for this BFO-poor SL. Indeed, that would be in agreement with the room temperature lattice parameters evolution versus x as shown in figure 1 where we identified a turning point of the structure at room temperature for x = 0.66. Below room temperature XRD measurements and analysis over a wider temperature range are needed to understand the BFO-poor SLs. A phase diagram is nevertheless proposed in figure 6 to summarize the above results and a comparison is made with previous results on BFO/LFO SLs [18].

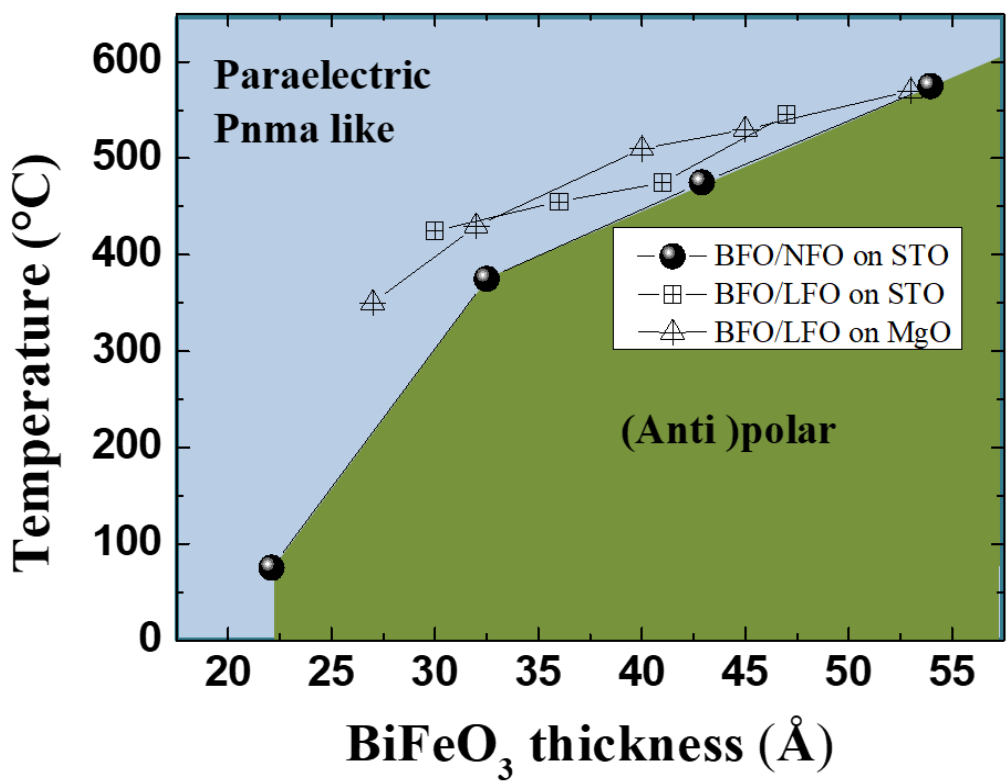

Figure 6. Temperature versus BiFeO$_3$ layer thickness diagram. Black spheres correspond to the T$_c$ observed within this report and comparison is provided with previous reports on similar BFO/LFO SLs (open symbols) [18]. The T$_c$ corresponding to the thinnest BFO layers (22.1 Å thick) deserves more attention to confirm the structural anomaly.

Figure 6 compares the previous work on BFO/LFO SLs [18] with the present results on BFO/NFO SLs. While in both systems the paraelectric to anti-polar $T_c$ scales with the BFO thickness we evidence lower stability of the anti-polar state in BFO/NFO for similar BFO thickness. The $T_c$ is indeed lower in BFO/NFO SLs compared to BFO/LFO SLs for similar BFO thickness. Because the available band gap values for both NFO and LFO in the literature are close while controversial, such behaviour is likely due to the higher structural competition at BFO/NFO heterointerfaces as suggested earlier, whereas the contribution of charge screening (effect on the depolarizing field) is not completely excluded. Indeed, interlayer strain effect based only on unit cell size may explain some structural behaviour but symmetry mismatch and electrostatic confinement need also to be taken into account to fully capture the physics in BFO/NFO SLs.

**Conclusion**

To conclude, for the first time, to the best of our knowledge, epitaxial BFO/NFO SLs were successfully grown by pulsed laser deposition and characterized by high-resolution X-ray diffraction and transmission electron microscopy. A room temperature structural change toward a paraelectric Pbnm orthorhombic phase is revealed on increasing the content of NFO in the period. A peculiar orthorhombic twin pattern is observed which propagates throughout the whole stacking and is believed to accommodate the interlayer strain and structural competition at heterointerfaces. HRTEM demonstrates a rich variety of twin patterns reminiscent of AFE-like state for x = 0.17 which according to previous reports may be better described by an AFE Pbnm state ($a^-a^-c^+/a^-a^-c^-$ rotation/tilt system) instead of the $PbZrO_3$ Pbam symmetry. We note the presence of sharp domain walls that similarly to $PbZrO_3$ may host peculiar properties (polar domain walls, antiphase boundaries). For x above 0.66 (the turning point), a typical paraelectric Pbnm structure is evidenced whatever the temperature while the temperature-dependent X-ray investigation demonstrates an anti-polar phase stability controlled by the BFO thickness. Comparison with similar BFO/LFO SLs [18] allows us to infer an influence of the strain mismatch over the range of anti-polar phase stability. Generalization to and confirmation with other rare-earth systems may help to better understand how the antiferroelectricity emerges and may pave the way to the design of efficient nanostructures for energy storage devices.

**Supplementary materials**

The supplementary materials contain complementary information about the layers growth depositions and the structural characterization of the individual layers.

**Acknowledgments**

The authors acknowledge Cabié Martiane for TEM sample preparation and Thomas Neisius for TEM observations. The Authors are grateful to Igor Lukyanchuk for illuminating discussions. This work is financially supported by the Conseil Régional des Hauts-de-France, the French National Research Agency (ANR) through the project EXPAND (ANR-17-CE24-0032), the Fonds Européen de Développement Régional (FEDER), and the European Union Horizon 2020 Research and Innovation actions MSCA-RISE-ENGIMA (No. 778072).

**DATA AVAILABILITY**

The data that support the findings of this study are available from the corresponding author upon reasonable request.

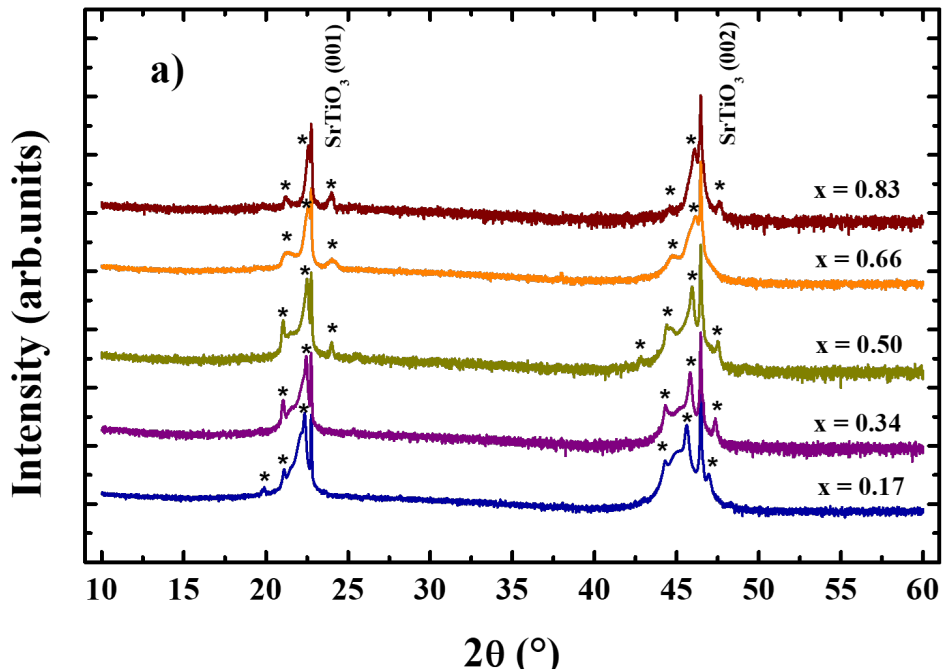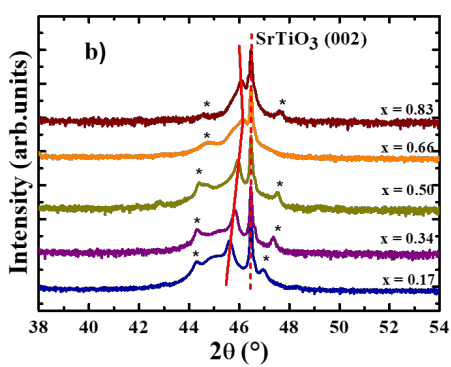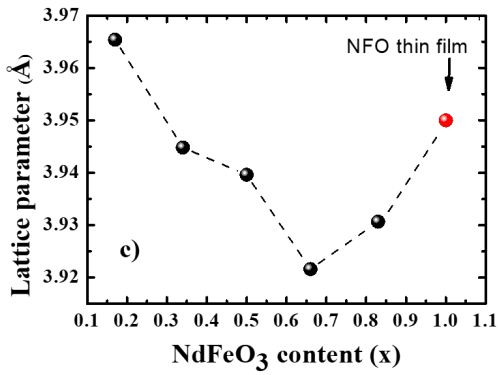

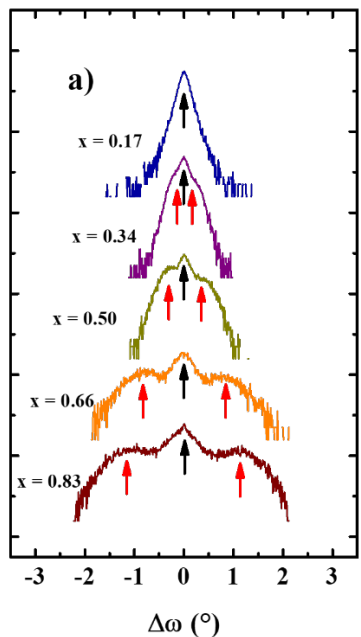
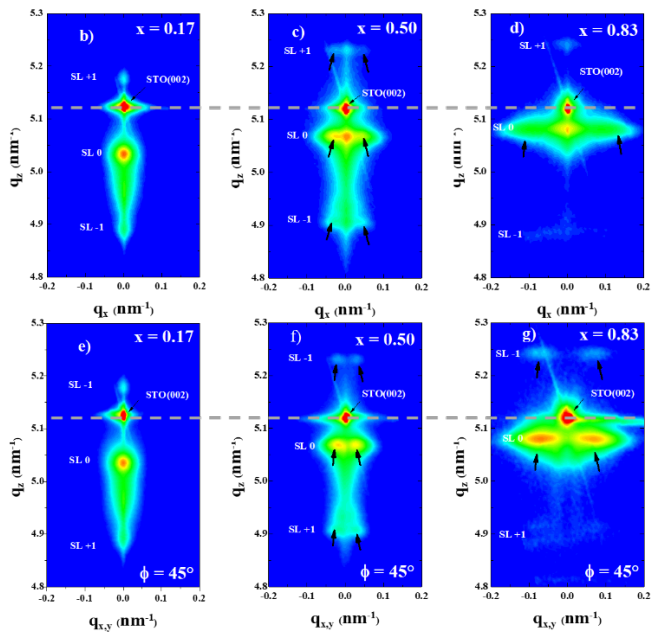
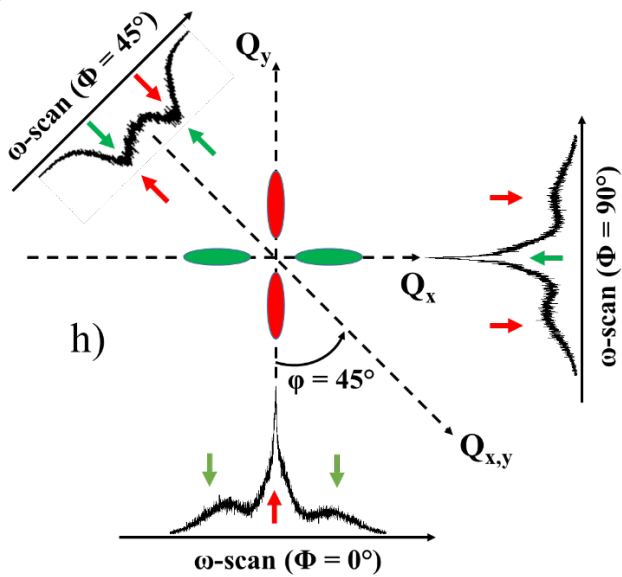

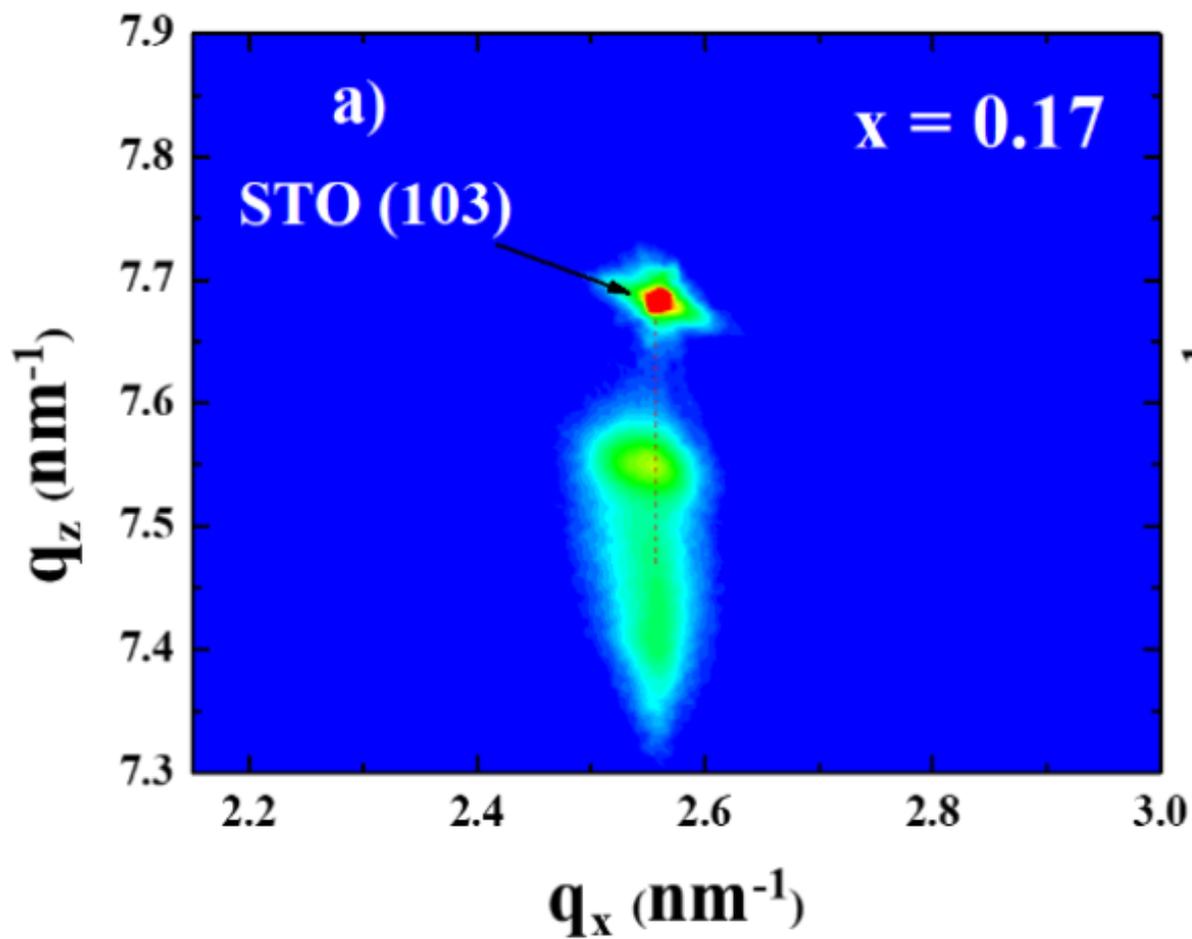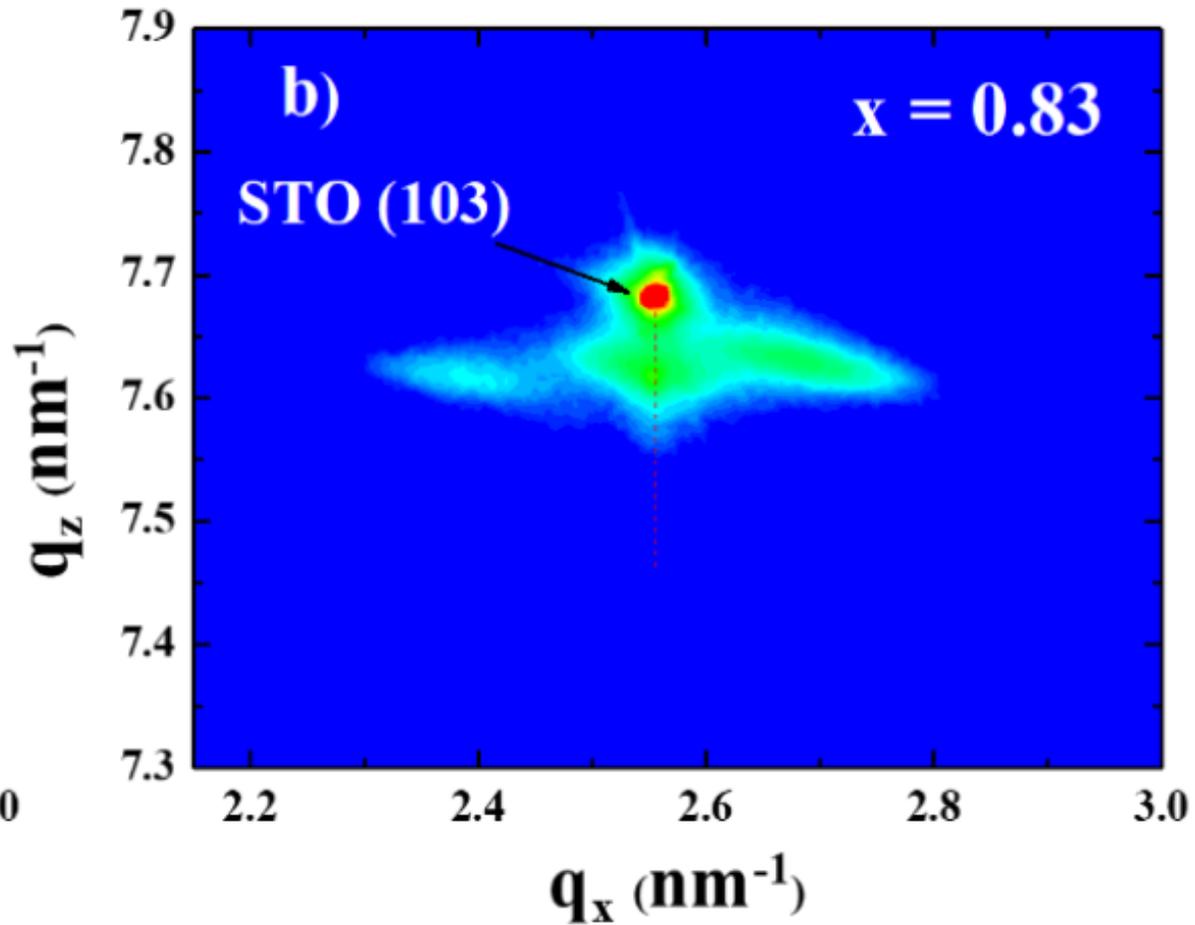

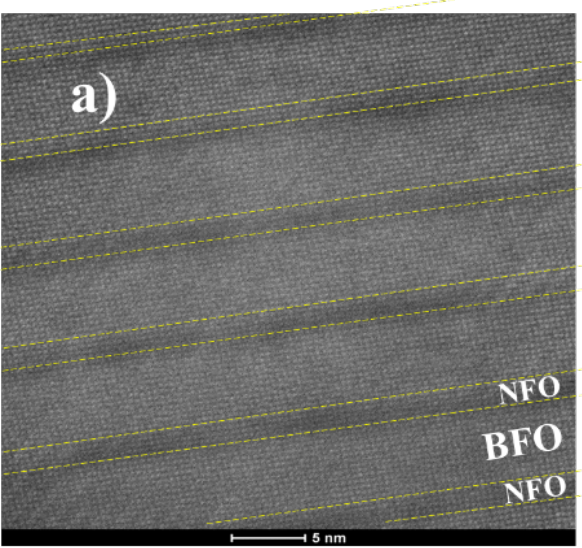
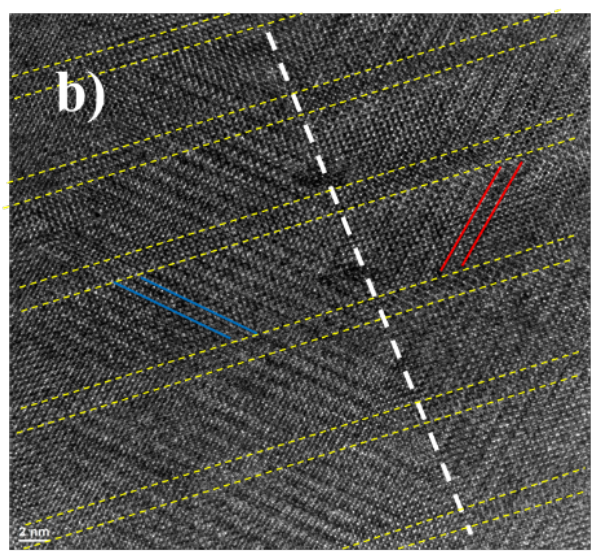
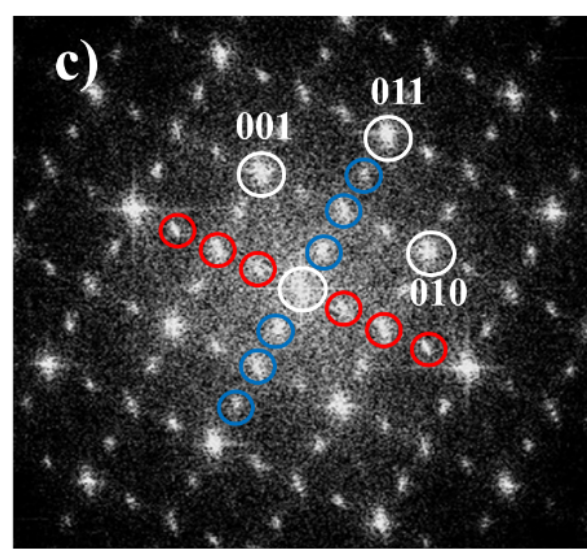
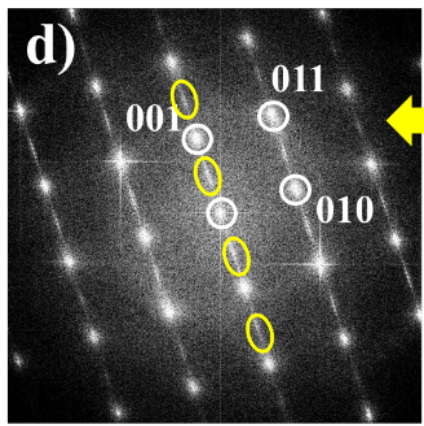
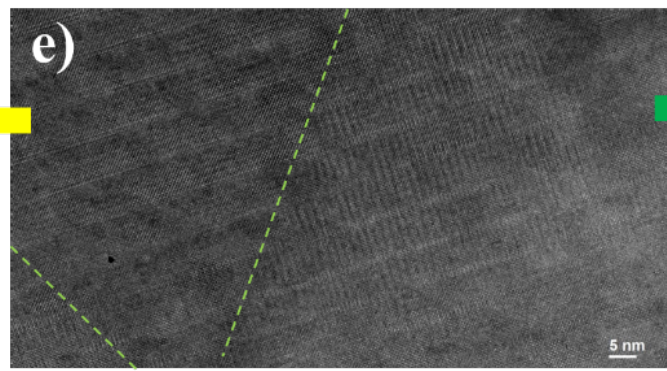
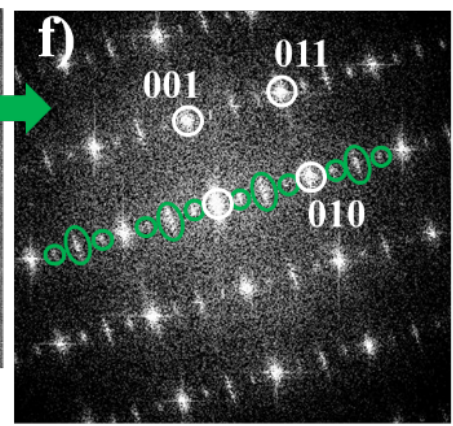

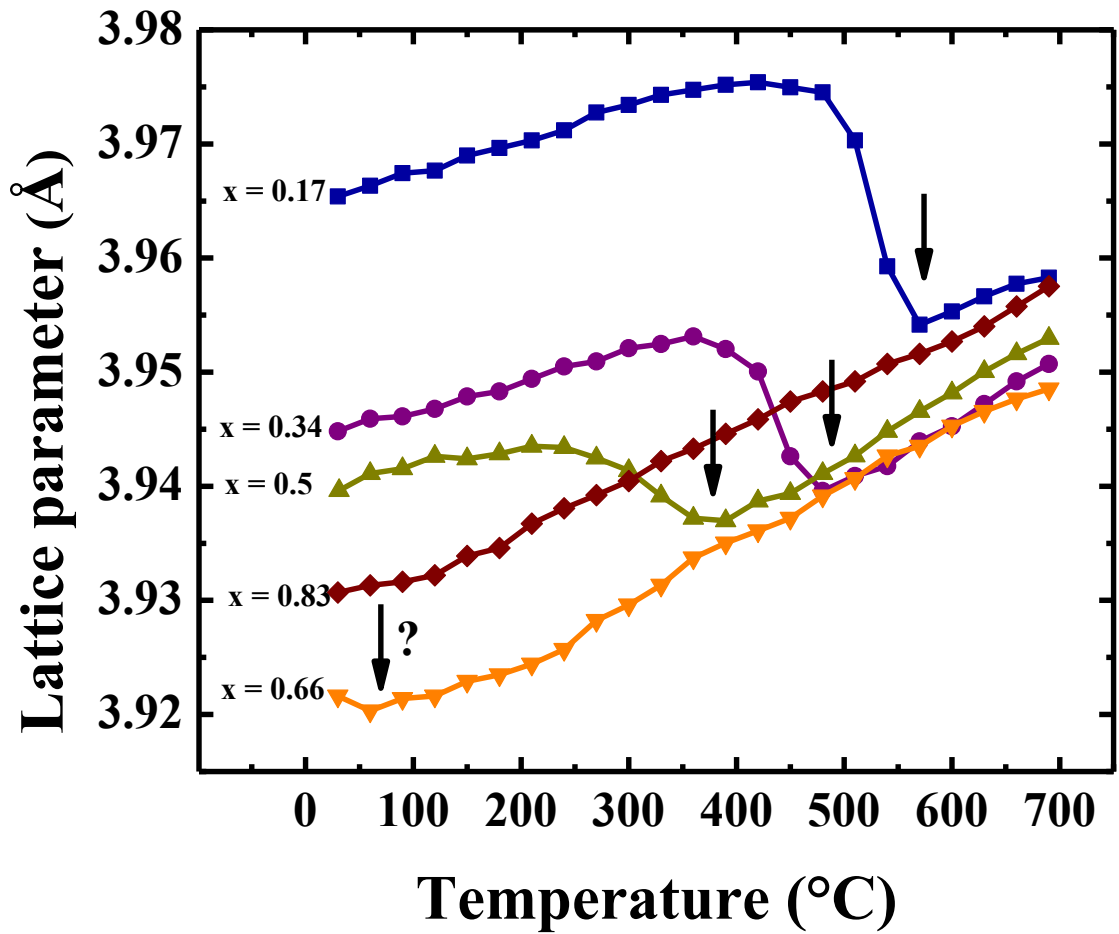

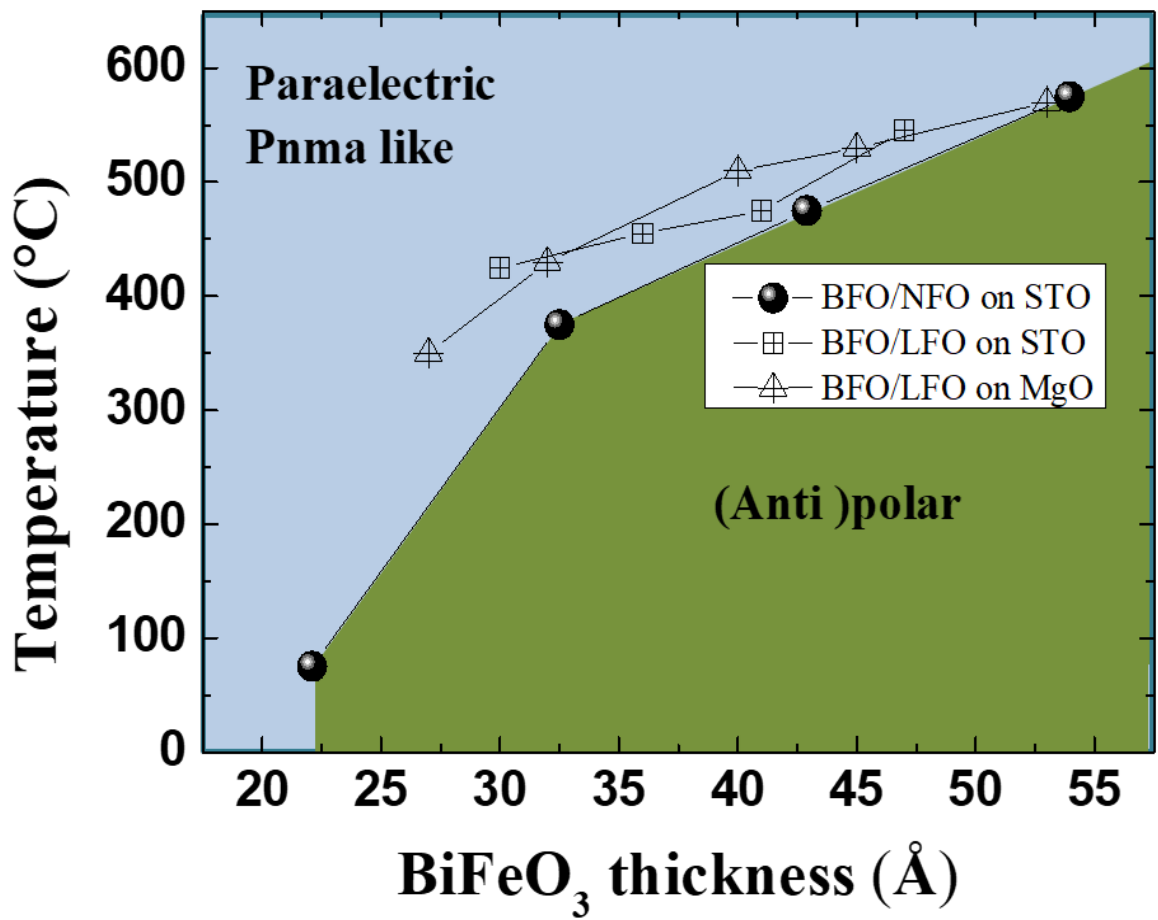